\documentstyle[preprint,prd,aps,floats]{revtex}
\input{psfig.tex}

\def\simlt{\mathrel{\spose{\lower 3pt\hbox{$\mathchar"218$}}
        \raise 2.0pt\hbox{$\mathchar"13C$}}}
\tighten
\begin{document} 
\draft
\preprint{
\begin{tabular}{rr}
Imperial/TP/95-96/23\\
astro-ph/9602040 \\
Submitted to {\em PRL}
\end{tabular}
}
\title{Satellite design parameters for detecting coherence in the
microwave sky}
\author{Andreas Albrecht and Benjamin D.\ Wandelt}
\address{Blackett Laboratory, Imperial College, Prince Consort Road
 London SW7 2BZ  U.K.}
\maketitle
\begin{abstract}
Recently it has been  realized that observations of
fluctuations in the cosmic microwave background (CMB) can reveal very
interesting information about the degree of coherence exhibited by the
perturbations at early times.  This fact should allow sufficiently
detailed observations to clearly
differentiate among several competing models of structure
formation.  We study the mission parameters required for a
satellite to address the issue of coherence. Our results emphasize the
importance of a small beamwidth, and support the cases for the PSI,
FIRE and COBRAS/SAMBA satellite proposals (in increasing order of
resolving power). Design parameters for a fourth proposal, MAP, have
not been made available.

\end{abstract}

\date{\today}


\renewcommand{\thefootnote}{\arabic{footnote}}
\setcounter{footnote}{0}

\section{Introduction}
 
Measurements of the Cosmic Microwave Background (CMB) anisotropies
have already had a tremendous impact on the field of cosmology. (For a
review see \cite{review}.) Future
measurements  promise to greatly increase this impact.  New
experiments will be measuring  the
anisotropies  with increasing resolution and accuracy, and our theoretical
understanding has advanced to the point where tremendous 
significance can be attached to these new data.

\begin{figure}[t]
\centerline{\psfig{file=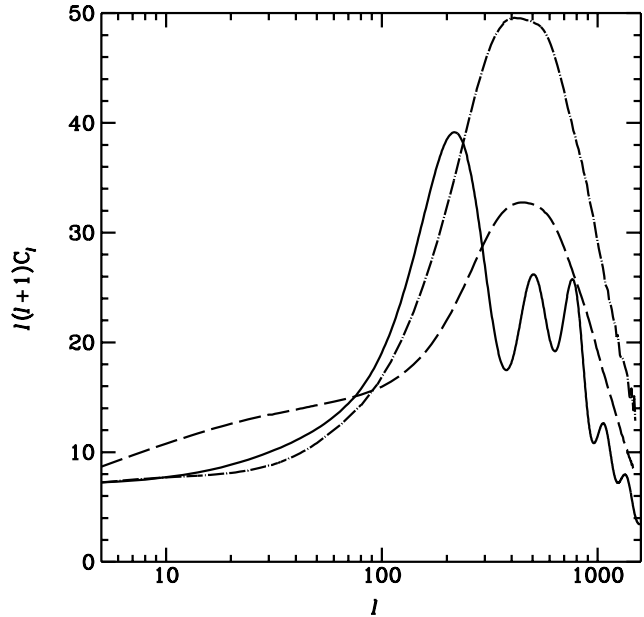,width=3.5in}}
\caption{ Angular power spectrum 
 of temperature fluctuations generated by
cosmic strings (dashed) and arising from 
a typical model of scale invariant
primordial fluctuations (solid) in arbitrary units. 
The shape of the string curve for $l \stackrel{<}{\sim} 100 $ (and
thus the height of the peak) is very sensitive to existing
uncertainties in string networks. This is illustrated by the second
string curve (dot-dashed) which represents a second plausible estimate
of the spectrum from a cosmic string model.
We show only the scalar contribution, in arbitrary units.  The string
curves are from Fig 4 in [3].}
\label{f4plus}
\end{figure}

Recently\cite{acfm,macf} it was realized that the degree of coherence in the
perturbations at early times will be reflected in striking features in
the angular 
power spectrum.  These features should allow observations to clearly
differentiate among several competing models of structure
formation. As illustrated in Fig \ref{f4plus}, the nature of 
the effect involves
the degree to which the so-called ``secondary Doppler peaks'' are
present in the angular power spectrum.  
It is worth noting that various types of perturbations 
(including certain ``decoherent active'' perturbations as 
discussed in \cite{macf} and \cite{ct}) can produce secondary peaks.
It is the presence or absence of these peaks that concerns us here.

The purpose of this {\it Letter} is to analyse the design parameters required
for a satellite to determine the presence or absence of secondary 
Doppler peaks.
We first describe the fundamental
methods and assumptions which are used throughout our discussion.
Then we step through a series of pairs of curves.  In each pair, 
one curve has ``standard'' secondary peaks and one does not.  
For each pair, we delimit the ``design parameter space'' which 
is able to resolve the differences between the two.  
 
\section{Methods}

\subsection{Modelling the satellite}

Our methods are the following.  We work entirely in ``$C_l$ space'',
where the $C_l$'s represent the angular power spectrum of microwave
anisotropies converted to an angular power spectrum using spherical
harmonics in the standard way. On a case-by-case basis we consider
pairs of curves 
$C_l$ which represent key differences one might wish to resolve.
The ``design parameters'' of the experiment are represented by an
effective ``beam width'' $\theta_{fwhm}$ and pixel noise $T_{Pix}$
which are converted into the expected variance in the $C_l$ for a
satellite
 of given fractional sky coverage $f$, using an extension of
the formula derived in \cite{Knox95}:
\begin{equation}
(\Delta C_l)^2=\frac{2}{f(2l+1)}
(C_l+4\pi\sigma^2_{pix}e^{l^2\sigma^2_b})^2
\label{eqn1}
\end{equation}
where $\sigma_{pix}=\frac{T_{pix}}{\sqrt{N_{pix}}}$; $N_{pix}$ is
the total number of pixels on the cut sky. Each pixel subtends a patch
of dimensions $\theta_{fwhm}\times\theta_{fwhm}$.
We
are assuming a Gaussian beamshape, and hence a beam function of the form 
$B_l=e^{-l(l+1)\sigma_b}$ \cite{White95}, where
$\sigma_b=\frac{\theta_{fwhm}}{2\sqrt{\ln{2}}}$. 
We model limited sky coverage by simply reducing the number of
independent modes per $l$ by the constant factor $f$ for all $l$. This
procedure is motivated in \cite{SSW94}, where the authors find the
naive ``guess'' $\sigma^2_{sample}\simeq (1/f)\sigma^2_{cosmic}$ to be
essentially correct.  We emphasise  that we view Eqn \ref{eqn1} as a
simple tool for generating realistic variances for the $C_l$'s as a
function of the parameters $\theta_{fwhm}$, $\sigma_{pix}$, and
$f$.
How these effective parameters are related to the fundamental design
parameters of a given satellite involves understanding specific
problems such as the process of foreground subtraction.  We
do not deal with these issues here.  

\subsection{Non-orthogonality of the $a_{lm}$'s}

Technically, the independent modes $a_{lm}$ which contribute to the
$C_l$'s are only orthogonal when considering the  
whole sky.  The orthogonal functions for a partial sky (besides being
fewer, as already discussed) will tend to get contributions from several
$a_{lm}$'s.  We expect this fact not to invalidate the rough arguments
below for the following reasons.  Firstly, a major factor in the whole
procedure is counting up the right number of independent modes. The
factor $f$ in Eqn \ref{eqn1} ensures that we are doing this correctly
(at least for large $l$). 
Secondly, for reasonably large sky coverage only a handful of nearby
$l$'s would be mixed to form the true orthogonal functions, and since
the $C_l$'s (and the error bars) are  not too rapidly varying with
$l$, we expect 
the error bars on the true orthogonal functions to be well estimated
by those of the related $C_l$'s.  Finally, although the first two
arguments probably break down for low values of $l$,  the results
presented here do not depend crucially on what is happening at low
$l$, so we expect overall not to be making any serious errors.

\subsection{Determining ``resolving power''}

We are trying to address the question: ``Given two theoretical 
curves f(l) and g(l), which part of
parameter space contains the experiments which can confidently
distinguish between them?''. To quantify this we calculate the
``distance'' $d$ between the two curves as
\begin{equation}
d=\sum^{l_{max}}_{l=2}(\frac{f-g}{\Delta C_l})^2
\end{equation}
using the errors as constructed
above. Under the assumption of near-Gaussianity, $d$ will be
nearly $\chi^2$-distributed and we can estimate the confidence $p$ with which
the curves can be distinguished for this particular experiment,
\begin{equation}
p=Q(\frac{l_{max}-1}{2},\frac{d}{2})
\end{equation}
where $Q$ is the incomplete Gamma Function.
We then find a sharp boundary in parameter space between a region of
nearly certain distinguishability and a region where experiments
fail to distinguish the two curves.  These are plotted in the top
panels  of Figs 2-5.  In regions above the heavy curve, there is
greater than $\%99.9$ probability of confusing the two curves, and below
the curve there is less than $\%0.1$ probability of confusion.  Note
that the $x$ coordinate in the top panel is noise (in $\mu K$) per unit
$\theta_{fwhm}$, where $\theta_{fwhm}$ (in arc minutes) is the $y$  
coordinate. 
In each of Figs 2-5 the lower panel shows
the two curves used, as well as the error bars associate with the
point indicated with a {$\times$} in the upper panel. (The $y$ axis is
in units of $\mu K^2$.)

\subsection{Further Discussion}

In this {\it Letter} we take the view that there are very strong reasons to
go for maximum sky coverage.  As well as investigating the low $l$
behaviour of the $C_l$'s, a single satellite will allow a wide range of
scales to be studied.  Even though individual ground based experiments can pin
down a rather narrow range of smaller angular scales, piecing several
such results together into a coherent picture can be tricky, not the
least because of the need to get the correct relative normalisation.
As we discuss below, the range of possible angular power spectra
is quite varied, and we regard the coherent picture which only a
satellite can produce to be essential to extracting clear cosmological
information from the CMB.  For example, in \cite{mh} Magueijo and
Hobson argue that
a long duration low sky coverage interferometry experiment would give
a simple way 
to detect the secondary peaks in Fig 2, but such an experiment would
be unable to differentiate the curves in Fig 5\footnote{J. Magueijo,
private communication.  Magueigo and Hobson also discuss single dish
experiments and find results broadly similar to ours.
}.

Throughout this {\it Letter}, we take the sky
fraction $f=0.4$. This seems to be a reasonable ``maximum'' sky
coverage which avoids excessive foreground contamination.
As a consequence of large sky coverage, one obtains a large number of
independent data points at small angular scales. This adds to the
ability of the satellite to resolve the secondary peaks, and is the origin of
the sharp contrast we find between the ``resolving'' and
``non-resolving regions''. 

\section{Exploring Theory Space}

\begin{figure}[t]
\centerline{\psfig{file=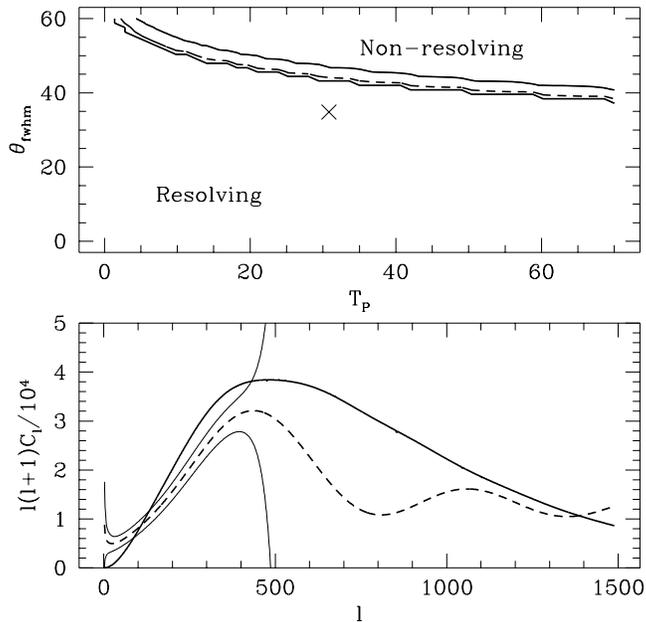,width=3.5in}}
\caption{ One of the string curves from Fig 1 paired with an
adiabatic (passive) model with $\Omega = 0.3$, $h = 60 $,$\Omega_b h^2
= 0.02$ and no tilt (in order to roughly match first peak positions).  A
wide range of satellites can distinguish between these two
curves. The technical details of the plot are provided in Section II} 
\label{c1}
\end{figure}

As we have already seen in Fig \ref{f4plus} a given low-coherence theoretical
model (such as cosmic 
strings) may well have an angular power spectrum which differs from
``similar'' high coherence model on {\em all} scales.  Figure \ref{c1} shows the
case of two similar curves, one with and one without secondary
oscillations.  Clearly these curves are sufficiently different to
allow a large (but still limited) range of experiments to distinguish
between them. 

However, we are faced with several theoretical uncertainties which
would allow coherent and decoherent model builders to fit the
data once it has come in.
Cosmic string models are relatively unconstrained at low
$l$. Even after these have been pinned down, there will be a wide
range of different types of ``decoherent'' models based, for example,
on different types of cosmic defects.
Uncertainties about the details of the inflaton potential will
continue to add flexibility to inflationary predictions. Both scenarios
are affected by the uncertainties in the values of cosmological
parameters such as $\Omega$, $\Omega_b$, $h$, and $\Lambda$, as well
as uncertainties in the ionization history. The
fact that the primary peak can occur in the same position for defect
and inflation models with different $\Omega$ generates a further
degeneracy. One of the exciting opportunities offered by the coherence
issue is that by searching for secondary Doppler peaks one can have an
impact on theoretical cosmology which transcends these uncertainties. 

\begin{figure}[t]
\centerline{\psfig{file=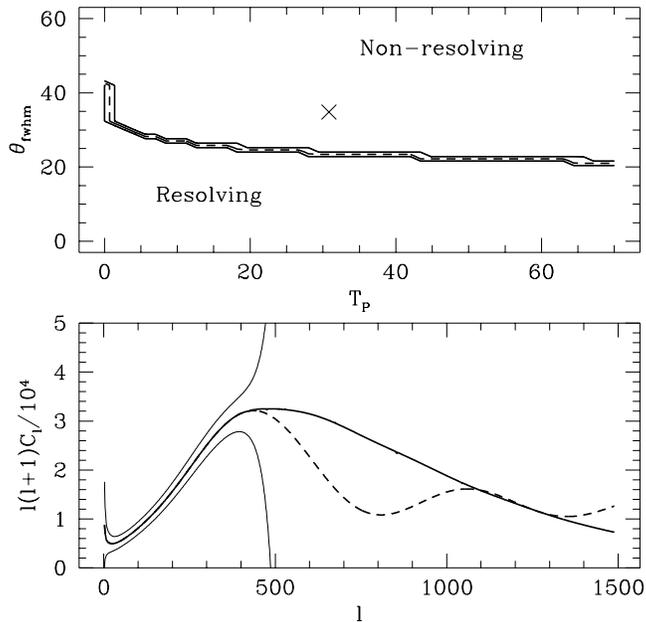,width=3.5in}}
\caption{The two curves from Fig 2 have been forced to match at low
$l$ (as one might expect theorists to do if this is where the data
lay).  Higher resolution experiments are required to pin down the
differences, which lie in the secondary Doppler peaks.} 
\label{c2}
\end{figure}

\begin{figure}[t]
\centerline{\psfig{file=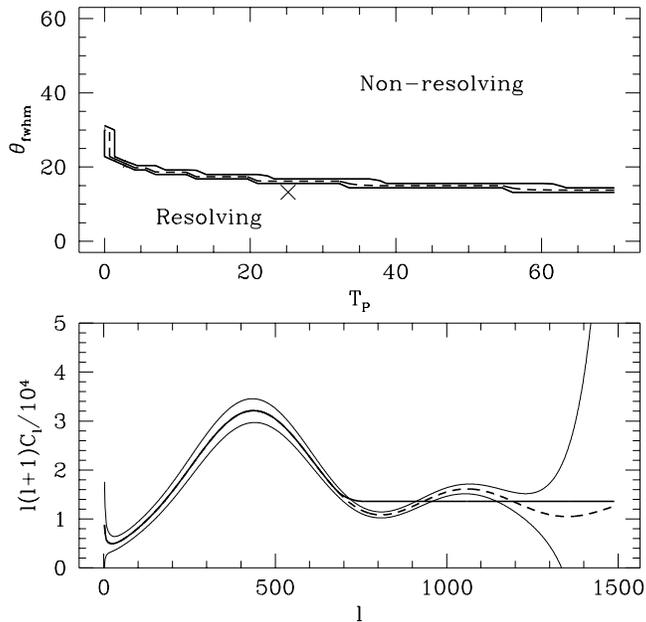,width=3.5in}}
\caption{Further modification of one curve from Fig 3 attempts to
maximize the similarity between the two curves without introducing
secondary peaks.  Even more pressure is placed on the satellite
design.} 
\label{c18}
\end{figure}

In what follows, we step through a series of pairs of curves, each
illustrating the kind of pressure that searching for secondary peaks
can put on satellite design. In each case, one of the curves has no
secondary peaks, and is forced to match the other (oscillating) curve
to varying degrees.  Many of these curves are constructed ``by hand'',
rather than being calculated from a particular detailed model.  We
feel that the prevailing theoretical uncertainties justify this approach.
In Fig \ref{c2} we have forced the two curves from Fig \ref{c1} to
match exactly at low $l$. Much higher resolution experiments are required
to resolve the difference between these two curves, since the
differences emerge only in the high $l$ region.
In Fig \ref{c18}, we have made a further modification of the curves
in Fig \ref{c2} so that the
curve with no secondary oscillations  tracks the inflation curve
as much as possible.  The useful region in
satellite parameter space is reduced even further.

\begin{figure}[t]
\centerline{\psfig{file=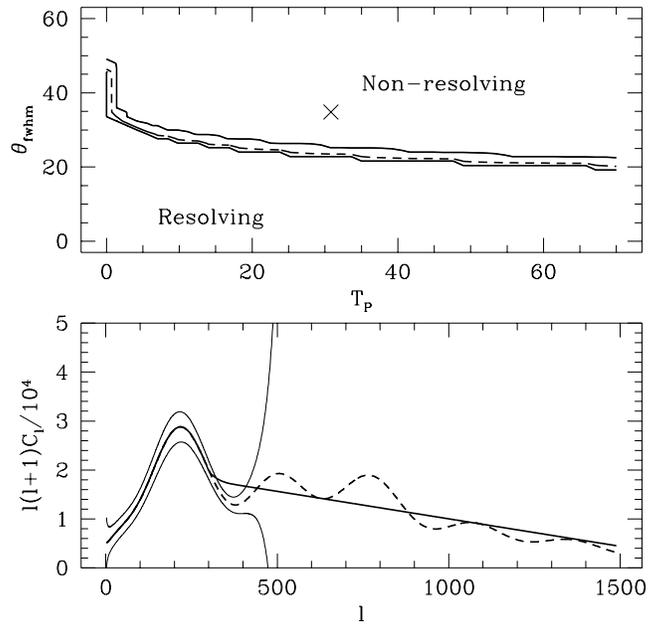,width=3.5in}}
\caption{A standard CDM curve is paired with a curve identical at low
$l$ and linear at high.  The selected error bars are from an
experiment which can clearly resolve the first peak, but which is
unable to confirm the presence of secondary oscillations.  Clearly the
search for secondary peaks places high demands on satellite design
even when the primary peak appears at larger angular scales.}
\label{c17}
\end{figure}

Cosmic string models tend to produce primary peaks at rather small
angular scales, so it is not surprising that looking for the secondary
oscillations makes high demands on the resolution.  Suppose however, that the
position of the primary peak is at the larger angular scale indicated
in the standard $\Omega = 1$ CDM model.  One would still want to
search for the presence of secondary peaks, as a test for
coherence\footnote{A standard CDM primary
peak {\em without} oscillations would be an extremely important
result.  It has been argued that this situation is very difficult to
achieve\protect\cite{macf} (see also \protect\cite{hw9602019}). 
}.  Figure \ref{c17} shows that even for this case the resolution
required to test for secondary oscillations is considerable.
Note that the experiment indicated by the $\times$ can easily resolve
the primary peak, but is unable to investigate the presence of
secondary oscillations.

\section{Conclusions}

The search for secondary oscillation in the CMB angular
power spectrum holds a deep significance for the field of cosmology.  We
have argued that a thorough search requires a satellite with
sufficient resolution.  The main point of this {\it Letter} is to make sure
that such a satellite gets built. There are four satellite proposals
presently under consideration and we outline their resolving power
here.  It is not
trivial to translate the realistic error bars (after eg foreground
subtraction) into the two parameters
used here\footnote{In order to allow a more systematic
discussion, we  are making our sets of curves (and other information)
available at ``http://euclid.tp.ph.ic.ac.uk/\~{ }albrecht/satellites/''
}.
  Still, \cite{jetal} suggests that the ``FIRE'' satellite  can be
roughly described by  
$\theta_{fwhm} = 8'$ and $T_{Pix} = 24.7\mu K$, and PSI by
$\theta_{fwhm}=12.6'$ and $T_{Pix} = 25\mu K$  Of these two, the lower
resolution error bars (PSI) 
appear on Fig \ref{c18} (the figure which places the greatest demands
on satellite design).  We conclude that both of these experiments
would be effective in searching for secondary oscillations. (The PSI
experiment is actually rather close to the edge of the ``resolving''
region.) The
COBRAS/SAMBA proposal has been analysed in \cite{te}.  Even the most pessimistic analysis from \cite{te} implies
an even greater resolving power than that of PSI or FIRE.  The design
parameters for the  fourth proposal, MAP\cite{map}, are not public
information. 

The presence or absence of secondary Doppler peaks in
the CMB power spectrum is of fundamental interest.  Our analysis
emphasizes small beamwidth as a crucial design parameter for their
detection, and thus
strengthens the case for a high resolution CMB satellite.

ACKNOWLEDGEMENTS: We thank P. Ferreira, J. Mageijo, D. Scott, and
M. White for helpful conversations, and we thank M. White for allowing
us to use his low $\Omega$ inflationary curves.  The standard CDM
curve is from N. Sugiyama\cite{ns}. We  acknowledge support from PPARC
(A.A.) and the Knowles Studentship of the University of London
(B.D.W.).

\pagebreak
\pagestyle{empty}

\end{document}